\documentclass[fleqn,usenatbib]{mnras}

\usepackage[T1]{fontenc}
\usepackage{ae,aecompl}

\usepackage{graphicx}	
\usepackage{amsmath}	
\usepackage{amssymb}	
\usepackage{newtxtext,newtxmath}

\newcommand{\nustar}{\textit{NuSTAR}}

\newcommand{\swift}{{\it Swift}}
\newcommand{\xmm}{{\it XMM-Newton}}

\newcommand{\red}{\textcolor{black}}
\newcommand{\ergs}{erg~cm$^{-2}$~s$^{-1}$}

\newcommand{\src}{UGC~11763}

\newcommand{\athena}{\textit{Athena}}

\title[Spectral Fitting of \src]{Comparing Reflection and Absorption Models for the Soft X-ray Variability in the NLS1 AGN \src}

\author[J. Jiang et al.]{
Jiachen Jiang,$^{1}$\thanks{E-mail: jcjiang12@outlook.com} Luigi C. Gallo,$^{2}$ Dirk Grupe,$^{3,4}$ Michael L. Parker$^{1}$
\\
$^{1}$Institute of Astronomy, University of Cambridge, Madingley Road, Cambridge CB3 0HA, UK\\
$^{2}$Department of Astronomy and Physics, Saint Mary's University, 923 Robie Street, Halifax, NS, B3H 3C3, Canada\\
$^{3}$\red{Department of Physics, Geology and Engineering Technology, Northern Kentucky University, Nunn Drive, Highland Heights, KY 41099, USA}\\
$^{4}$Department of Physics, Earth Science, and Space System Engineering, Morehead State University, 235 Martindale Dr, Morehead, KY 40351, USA\\
}

\date{Accepted XXX. Received YYY; in original form ZZZ}

\pubyear{2022}

\begin{document}
\label{firstpage}
\pagerange{\pageref{firstpage}--\pageref{lastpage}}
\maketitle

\begin{abstract}
We present a spectral analysis of two \xmm\ observations of the narrow-line Seyfert 1 galaxy \src. \src\ shows very different soft X-ray spectral shapes in the two observations separated by 12 years. Three spectral models are considered to explain the multi-epoch X-ray variability of \src, \red{one based on the relativistic disc reflection model, one based on multiple partially-covering absorbers combined with the warm corona model}, and a hybrid model. In the first model, the X-ray variability of \src\ is caused by the emission from a compact coronal region with a variable size. The resulting disc reflection component changes accordingly. A warm absorption model with a modest column density is required in this model too. In the \red{partially-covering} absorption scenario, the X-ray variability of \src\ is caused by the variable covering factors of two absorbers located within a region of $r<\approx100r_{\rm g}$. Moreover, the temperature and strength of the warm corona have to change significantly too to explain the variable underlying soft X-ray emission. Lastly, we investigate the possibility of variable \red{intrinsic} power-law emission from the hot corona combined with variable absorption in \src\ \red{without changing the geometry of the corona} in the third model. This hybrid model provides a slightly better fit than the \red{partially-covering absorption} model with improvements in fitting the iron emission band. Current CCD-resolution data cannot distinguish these spectral models for \src. Future high-resolution X-ray missions, e.g. \athena\ and \textit{XRISM}, will test them by resolving different spectral components.
\end{abstract}




\section{Introduction} \label{intro}

Narrow-line Seyfert 1 galaxies (NLS1s) are a class of peculiar Seyfert 1 galaxies (Sy1s) with strong Fe~\textsc{ii} emission, narrow H$\beta$ emission and weak [O~\textsc{iii}] emission in the optical band in comparison with other Sy1s \citep{goodrich89}. It is believed that NLS1s host low-mass supermassive black holes (BHs) that are accreting near or around the Eddington limit \citep[e.g.][]{boroson02, grupe04a}, although some studies point out that the BH mass measurements of NLS1s might be biased towards low values when radiation pressure is ignored in the broad line region model for the H$\beta$ emission of NLS1s \citep{marconi08}.

In the X-ray band, the nuclei of NLS1s often show rapid, large-amplitude X-ray variability \citep[e.g.][]{mchardy95, boller03, smith07, jin17, alston19} and steep X-ray continuum emission \citep{puchnarewicz92,boller96,grupe98}. \citet{gallo06} finds that NLS1s show complex spectral features in the X-ray band, especially during their low flux state. In particular, NLS1s often show strong excess emission below 2--3\,keV in addition to the hard X-ray continuum extrapolated to the soft X-ray band \citep{boller96, piconcelli05}.

Different models have been proposed to explain the broad-band X-ray spectra of NLS1s \citep[see a review of this topic in][]{gallo18}. One of them is the disc reflection model \citep[e.g.][]{fabian04,larsson08, miniutti09, brenneman11,reis12, tan12, risaliti13, gallo13, walton13, parker14,marinucci14, jiang19,jiang20b}. The surface of the optically thick disc produces reprocessed emission of the illuminating coronal emission in the soft X-ray band and the Compton-scattering hump in the hard X-ray band. They are often referred to as the disc `reflection' component \citep{fabian89}. The most prominent feature of a disc reflection component is the Fe~K$\alpha$ emission line around 6.4\,keV, which is broadened by relativistic effects in the vicinity of the BH \citep{tanaka95}. The disc reflection model of Seyfert active galactic nuclei (AGNs) is also supported by the discoveries of X-ray reverberation lags in the soft X-ray band \citep[e.g.][]{fabian09,demarco13}, the Fe~K band \citep[e.g.][]{kara16} and the hard X-ray band \citep[e.g.][]{zoghbi14,kara15}

Another model used to explain X-ray spectra of Sy1s is the double partially-covering absorption model \citep{tanaka04}. In this model, multiple high-column density absorbers crosses our line of sight towards the X-ray emission region. The absorbers require a tricky geometry to partially cover the compact X-ray emission region as suggested by X-ray data \citep{reynolds09}. In this model, the absorbers produce strong Fe~K edge in the spectrum, which is used to explain the steep spectra of some NLS1s \citep[e.g.][]{gallo15}. 

An additional component is still required to fit the soft excess emission when one uses the absorption model to fit the data in the Fe~K band. For instance, a soft power-law model was used to explain the soft excess emission of the NLS1 1H~0707$-$495 in combination with the absorption model in \citet{tanaka04}. Such a soft power-law component is proposed to originate in a warm coronal region with a high optical depth of $\tau_{\rm T}=10-20$ and a low temperature of $kT_{\rm e}<1$\,keV \citep{magdziarz98, petrucci01, czerny03, jin17, ursini20}. At such a low temperature, atomic opacity dominates over the Thomson opacity. Strong emission and absorption features could be shown in the spectrum of the warm corona in contradiction to the data of Sy1s \citep{garcia18}. Simulations by \citet{petrucci20}, however, suggest that the warm corona has to be heated by hard X-ray continuum from the hot corona and the disc emission from below, where the ions in the upper layer of the warm corona can be completely ionised showing no or weak emission and absorption lines. Photoionisation models also find that a higher accretion rate in the accretion disc leads to a warm corona producing stronger soft excess emission, similar to what is suggested by the X-ray data of Sy1s \citep{ballantyne20}.

In this work, we investigate how the models introduced above may explain the \xmm\ observations of the \red{active galaxy} \src\ at $z=0.063$ \citep[23 32 27.8, +10 08 19, ][]{clements81, huchra99}. \red{\src\ was classified as a NLS1 galaxy by \citet{boroson92,constantin03}. The H$\beta$ emission of this NLS1 has a width of 2250--2800 km\,s$^{-1}$ \citep{boroson92,grupe04,mullaney08}, which is higher than the values of typical NLS1s.} The mass of the central BH in \src\ is estimated to be around $4.57\times10^{8}M_{\odot}$ \citep{peterson04} \red{using  broad emission-line reverberation-mapping data. \citet{ho08} further lowered this mass measurement by a factor of 1.8 to $2.5\times10^{8}M_{\odot}$ for consistency with the virial mass zero point\footnote{\red{See Footnote 4 in \citet{greene05}.}} adopted by \citep{greene05}.} These measurements are near the upper limit of the BH mass distribution of NLS1s \citep{grupe04}. \citet{peterson04} estimated the luminosity of \src\ at 5100\AA\ to be $\log(\lambda L_{\lambda})=44.46\pm0.04$. Assuming $L_{\rm Bol}=9\lambda L_{\lambda}$ \citep{kaspi00} and a BH mass of $4.6\times10^{8}$\,$M_{\odot}$ \citep{peterson04}, the Eddington ratio of \src\ is around 5\%.

In the X-ray band, \citet{cardaci09} reported a steep X-ray emission from \src\ and strong soft excess emission as well as the \textit{EXOSAT} data of the same source \citep{singh91}. \citet{cardaci09} also found Fe `Unresolved Transition Array' (UTA) absorption in the soft X-ray band of \src, which indicates the existence of photoionised absorbers, e.g. warm absorption commonly seen in AGNs \citep[e.g.][]{reynolds97,george98}. Large-amplitude X-ray variability of \src\ has been realised for decades and was first discovered by \citet{singh91,grupe01}. For instance, the X-ray flux of this source measured by \textit{ROSAT} varies by 50\% in the 0.1--2\,keV band in only one year.

In this paper, we present spectral analysis of the \xmm\ data of \src\ using two archival observations. We consider both the reflection-based and the absorption-based model. In Section \ref{data}, we introduce our data reduction processes; in Section \ref{analysis}, we present three models for the spectra of \src; in Section \ref{discuss}, we discuss our results; in Section \ref{conclude}, we conclude our work. 

\section{Data Reduction} \label{data}

\begin{table*}
    \centering
        \caption{The list of \xmm\ observations analysed in this work. $F_{\rm 0.3-3keV}$ and $F_{\rm 3-10keV}$ are the observed flux of \src\ measured by pn in corresponding energy bands.}
    \label{tab_obs}
    \begin{tabular}{cccccc}
    \hline\hline
    Name & Obs ID & Date & Time & $\log(F_{\rm 0.3-3keV})$ & $\log(F_{\rm 3-10keV})$\\
     &   &  & ks & \ergs & \ergs \\
    \hline
    obs1 & 0150470701  & 2003-05-16 & 38 & $-11.467\pm0.002$ & $-11.535^{+0.009}_{-0.007}$ \\
    obs2 & 0744370201  & 2015-05-01 & 33 & $-11.141\pm0.005$ & $-11.483\pm0.019$\\
    \hline
    \end{tabular}
\end{table*}

\begin{figure}
    \centering
    \includegraphics[width=\columnwidth]{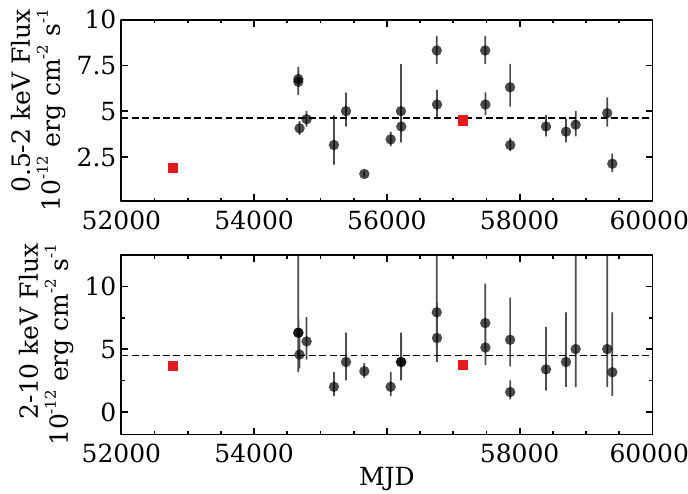}
    \caption{Long-term X-ray lightcurves of \src\ in units of $10^{-12}$\,\ergs\ (top: 0.5--2\,keV; bottom: 2--10\,keV; black circles: \swift\ XRT observed flux; red squares: \xmm\ pn observed flux). The error bars of the red squares are smaller than the sizes of points. The dashed lines show the average flux of \src\ in the two energy bands.}
    \label{pic_swift}
\end{figure}

We use the European Photon Imaging Camera (EPIC) observations for X-ray continuum modelling. A full list of observations used in our work is in Table\,\ref{tab_obs}.

The EPIC data are reduced using \red{V20.0} of the \xmm\ Science Analysis System (SAS) software package. The version of the calibration files  is v.\red{20220407}. We first generate a clean event file by running EMPROC (for EPIC-MOS data) and EPPROC (for EPIC-pn data). Then, we select good time intervals by filtering out the intervals that are dominated by flaring particle background. These high-background intervals are where the single event (PATTERN=0) count rate in the >10~keV band is larger than 0.35~counts~s$^{-1}$ (0.4 counts~s$^{-1}$) for MOS (pn) data. By running the EVSELECT task, we select single and double events for EPIC-MOS (PATTERN<=12) and EPIC-pn (PATTERN<=4, FLAG==0) source event lists from a circular source region of 35 arcsec. Background spectra are extracted from a nearby circular region of 60 arcsec. No obvious evidence of pile-up effects has been found in obs1 in 2003. The pn and MOS instruments were operated in the full frame mode in 2015. Some evidence of pile-up effects were found. An annulus region with an inner radius of 10 arcsec and an outer radius of 35 arcsec is used to extract source spectra. The inner radius is chosen according to the EPATPLOT tool in SAS\footnote{\red{We use EPATPLOT to estimate the full-band observed-to-model ratios for single and double events based on expected pattern distribution functions from the latest calibration data. This ratio is $0.96\pm0.01$ for singles and $1.12\pm0.02$ for doubles when a circular region is used to extract source products from the pn observation, suggesting evidence of pile-up. Singles  and doubles ratios are respectively $0.99\pm0.02$ and $1.03\pm0.04$ when an annulus region with an inner radius of 10 arcsec is used. Similar conclusions are found for MOS observations.}}. Last, we create redistribution matrix files and ancillary response files by running RMFGEN and ARFGEN. In this work, we consider the full 0.3--10\,keV band of EPIC data. The spectra are grouped to have a minimum of 20 counts per bin and oversample by a factor of 3.

\section{Soft X-ray Variability of \src}

\begin{figure*}
    \centering
    \includegraphics[width=\textwidth]{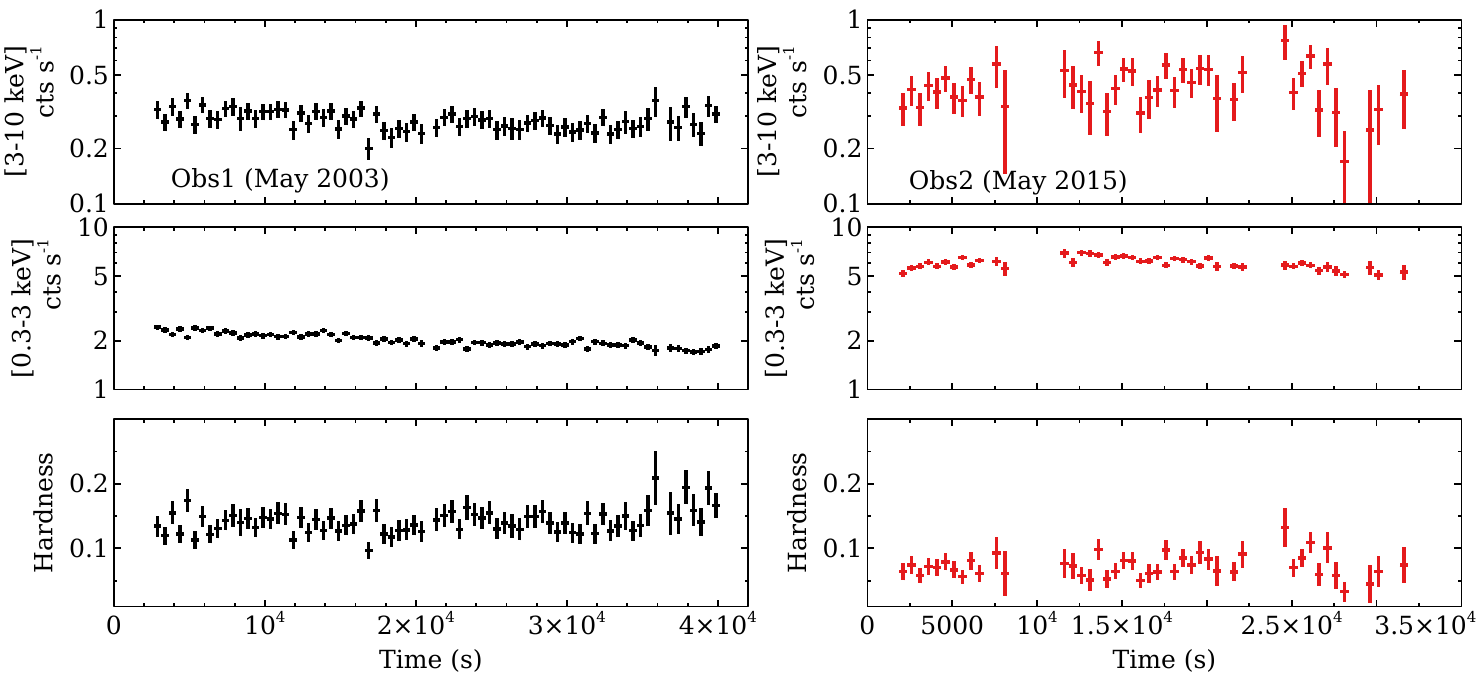}
    \caption{\xmm\ pn lightcurves of \src\ in the 3--10\,keV and 0.3--3\,keV bands (left: obs1; right: obs2). Note that the y-axes are in log scale. The bottom panels show the hardness ratio defined as the count rate ratio between 3--10\,keV and 0.3--3\,keV bands. Unlike typical NLS1s, \src\ does not show extreme X-ray variability on timescales of kiloseconds.}
    \label{pic_xmm_lc}
\end{figure*}

Fig.\,\ref{pic_swift} shows the long-term X-ray lightcurves of \src\ in the 0.5--2\,keV and 2--10\,keV bands. In particular, the \xmm\ observations analysed in this work are marked with the red squares in the figure. The X-ray flux of \src\ shows variability on timescales of months and years . For instance, the 0.5--2\,keV flux of this source increased from the reported minimum flux of $1.6\times10^{-12}$\,\ergs\ in April 2011 to the maximum flux $8.3\times10^{-12}$\,\ergs\ in April 2014. X-ray variability with a similar amplitude was realised in previous observations \citep{singh91,grupe01}.

In this work, we focus on the \xmm\ observations of this source. The first observation was taken during a low flux state and the second was taken during a higher flux state in the soft X-ray band. In comparison with the soft X-ray band, the 2--10\,keV flux of \src\ does not show large amplitude variability. This leads to the question--what may cause the X-ray variability of \src\ confined below 2\,keV.

We show EPIC-pn lightcurves extracted from the two \xmm\ observations in Fig.\,\ref{pic_xmm_lc}. Unlike other extreme NLS1s \citep[e.g.][]{mchardy95, boller03, smith07, jin17, alston19}, \src\ does not show rapid and large-amplitude variability on timescales of kiloseconds in the X-ray band. 
The lack of rapid variability on kilosecond timescales and the evidence of large-amplitude variability on longer timescales might be related to the relatively higher BH mass of \src\ \citep[a few times $10^{8}M_{\odot}$,][]{peterson04, ho08} in comparison with other NLS1s.

\section{Spectral Analysis} \label{analysis}


\begin{figure}
    \centering
    \includegraphics[width=\columnwidth]{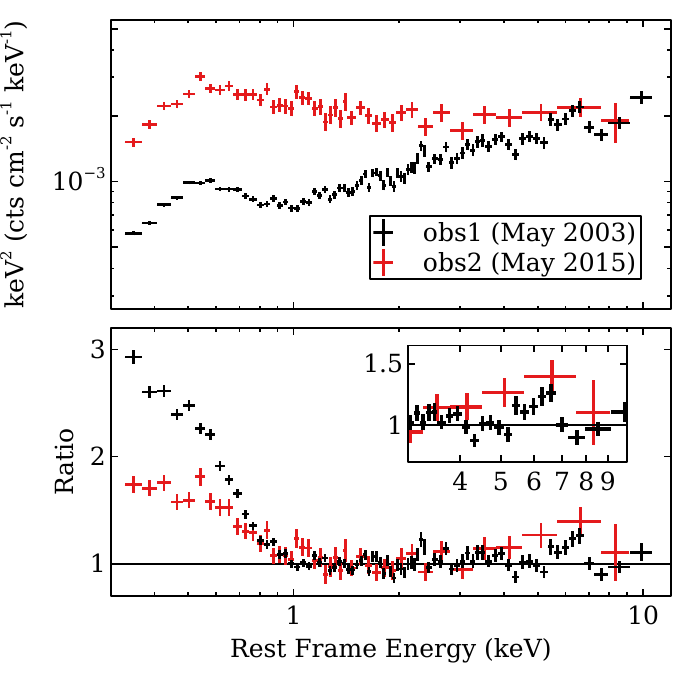}
    \caption{Top: folded EPIC-pn spectra of \src\ corrected for effective area.  Two observations, separate by 12 years, show significantly different soft X-ray emission below 1\,keV. Bottom: data/model ratio plots of the pn spectra of \src\ using the best-fit absorbed power-law model for the 1--10\,keV band of each observation. The smaller panel zooms in on the 4--9\,keV band. Both observations show evidence of Fe~K$\alpha$ emissions.}
    \label{pic_pl}
\end{figure}

We use the XSPEC software (v.12.12.1) for spectral analysis \citep{arnaud96}. We start our analysis by fitting the spectra above 2\,keV with an absorbed power law. The \texttt{tbnew} model is used to account for Galactic absorption \citep{wilms00}, which is estimated to be $N_{\rm H}=4.6\times10^{20}$\,cm$^{-2}$ \citep{willingale13}. Corresponding data/model ratio plots are shown in Fig.\,\ref{pic_pl}. A zoom-in of the 3--10\,keV band is shown in the lower panel. Both epochs show evidence of Fe~K$\alpha$ emission in the Fe~K band. Meanwhile, two epochs show different spectral shape in the soft X-ray band. 

In the rest of the section, we focus on modelling the spectra of \src\ using the absorption, disc reflection and hybrid models.

\subsection{Relativistic Disc Reflection Model (\red{Model 1})}

\subsubsection{Model set-up}

In the disc reflection scenario, the soft excess is interpreted as part of the reprocessed emission from the inner accretion disc \citep[e.g.][]{crummy07,jiang20b}. To model the disc reflection component, we use the \texttt{relxilld} model \citep[\red{Model 1},][]{garcia16}. A broken power-law emissivity profile parameterised by q1, q2 and $R_{\rm b}$ is considered. Other free parameters include the spin of the BH, the inclination angle, the iron abundance and the electron number density of the disc. The reflection fraction parameter of \texttt{relxilld} is set to be a free, positive number, so the model returns both the hot coronal emission and the corresponding disc reflection spectrum.

\begin{figure}
    \centering
    \includegraphics[width=\columnwidth]{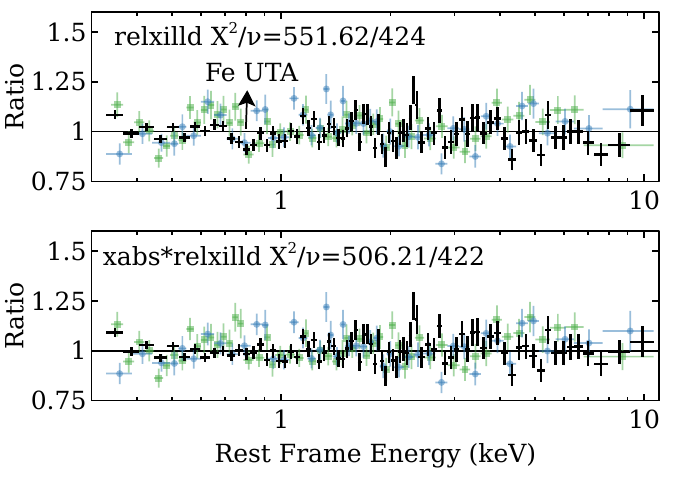}
    \caption{Data/model ratio of the EPIC pn (black), MOS1 (blue crosses) and MOS2 (green squares) spectra of \src\ in 2003 (obs1) using \texttt{relxilld} (top). Absorption features at 0.8\,keV (16--17\AA) are shown in the data. They are a blend of Fe M-shell absorption lines (Fe `Unresolved Transition Array'), suggesting the existence of ionised absorbing gas in \src. Similar results were found in the grating data of \src\ \citep{cardaci09}. An additional low-ionisation absorption model improves the fit by $\Delta\chi^{2}=46$ with two more free parameters (bottom).}
    \label{pic_warm}
\end{figure}

By applying \texttt{relxilld} to obs1, we find evidence of absorption features at 0.8\,keV. See Fig.\,\ref{pic_warm} for corresponding data/model ratio plots. The goodness of this fit is $\chi^{2}/\nu=$\red{551.62/424}. The absorption features correspond to Fe UTA, suggesting the existence of photoionised absorbers in a low ionisation state, e.g. warm absorbers commonly seen in AGNs \citep[e.g.][]{lee01,ebrero16}. \citet{cardaci09} pointed out similar residuals in the spectra of \src\ extracted from obs1 when applying a simple model including Galactic absorption, a power law, a black body and Fe~K$\alpha$ emission line. 

To fit the absorption features, we use a tabulated version of the \texttt{xabs} photoionised absorption model \citep{steenbrugge03} from SPEX \citep{kaastra96}, implemented as an XSPEC table model by \citet{parker19} and available from www.michaelparker.space/xspec\_models. The version used here assumes the photoionisation is driven by a $\Gamma=2$ power-law input spectrum, and covers ionisations from $\log(\xi)=-4$ to 5, with parameters for the column density, velocity broadening, and covering fraction.



When fitting the warm absorption features in the data, we assume low velocity broadening of 100\,km\,s$^{-1}$ and a full-covering geometry. The assumption for this geometry is based on the long distance between warm absorbers and the X-ray emission region: typical warm absorbers are estimated to be near the broad line region \citep[e.g.][]{reynolds95}. 

By including the \texttt{xabs} model, the fit is significantly improved, e.g. $\Delta\chi^{2}=45$ and two more free parameters. We, therefore, conclude that our best-fit model is \texttt{tbnew * xabs * relxilld} (\red{Model 1}). We only obtain an upper limit of the Galactic column density ($N_{\rm H}<6\times10^{20}$\,cm$^{-2}$). This parameter is thus fixed at the nominal value calculated by \citet{willingale13} ($4.6\times10^{20}$\,cm$^{-2}$). Similar conclusions are found for obs2.

\subsubsection{Multi-epoch analysis} \label{ref_multi}

\red{We fit the two observations of \src\ simultaneously with Model 1 to better understand the spectral variability in this object.} Best-fit parameters are shown in the first two columns of Table\,\ref{tab_ref}.

Some of the parameters for the two observations are consistent within their uncertainty ranges. For instance, we only obtain an upper limit of the density of the reflection surface of the disk for each observation. Both values are consistent with a low density of $10^{15}$\,cm$^{-3}$, which was commonly assumed in the disc reflection modelling of AGN data \citep[e.g.][]{ross05}. In addition, the spin of the central BH, the inclination angle and iron abundances of the accretion disc are not expected to change on observable timescales. We obtain consistent measurements of these parameters for two observations, which increases our confidence in the choice of our reflection model. Lastly, the ionisation state of the warm absorber also remains consistent in these two observations. 

\begin{figure}
    \centering
    \includegraphics[width=\columnwidth]{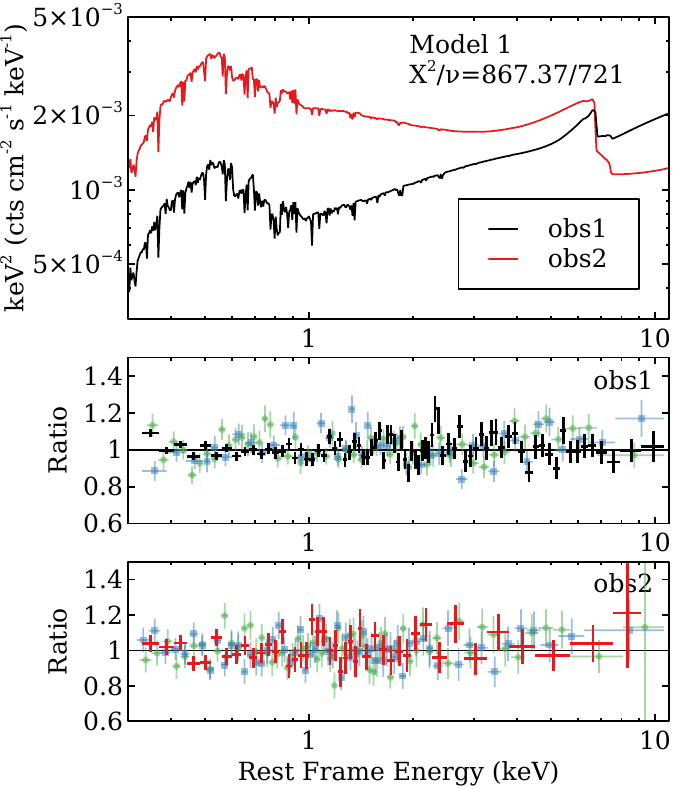}
    \caption{Best-fit models for two \xmm\ observations using Model 1 and corresponding data/model ratio plots (red: obs1 pn; black: obs2 pn; blue squares: MOS1; green circles: MOS2). Spectra extracted from two observations are fit together. See text for more details.}
    \label{pic_ref}
\end{figure}

We conduct multi-epoch spectral analysis with all the parameters mentioned above linked between two observations. By doing so, we obtain a good fit for both observations with $\chi^{2}/\nu=$\red{867.37/721}. The best-fit parameters are shown in Table\,\ref{tab_ref} and the best-fit models are shown in the upper panel of Fig.\,\ref{pic_ref}. Corresponding data/model ratio plots are shown in the lower panels of Fig.\,\ref{pic_ref}. 

\subsubsection{Results} \label{ref_measure}

By comparing the best-fit parameters of \red{Model 1} for the two \xmm\ observations of \src, we find that the coronal emission shows a softer-when-brighter pattern--the photon index of the coronal emission increases from 2.26 to 2.41. This is commonly seen in other AGN too \citep[e.g.][]{jiang18,wu20}.

The fit of the disc reflection spectrum in obs2 requires a flatter disc emissivity profile, i.e. a lower q1 and a higher $R_{\rm b}$, suggesting a change in the coronal geometry, e.g. a more extended corona in \src\ during obs2 than obs1 \citep{dauser13}. The anti-correlation between the X-ray flux and the reflection fraction parameter suggests a similar conclusion. When the coronal region is more compact, more coronal photons are lost to the event horizon. On the other hand, disc reflected photons from a more extended emission region is less affected by light-bending effects, resulting in a higher reflection fraction in the spectrum \citep[e.g.][]{miniutti03}. 


In summary, the spectral difference between obs1 and obs2 is explained by the variable coronal emission in \red{Model 1}. The disc reflection component changes accordingly. During the low flux state (obs1), the coronal region is more compact as suggested by the steeper emissivity profile and the higher reflection fraction parameter. Meanwhile, the variable column density of the warm absorber also contributes to the soft X-ray variability while the ionisation state of the warm absorber remains consistent. 

The spin of the BH, the inclination angle of the inner disc and the iron abundances of the disc are not expected to change on observable timescales. By linking these parameters of two observations, we obtain $i=32\pm2^{\circ}$, $a_{*}>0.97$ and $Z_{\rm Fe}=4.8\pm1.2Z_{\odot}$.

\begin{table*}
    \centering
    \begin{tabular}{ccccccc}
    \hline\hline
         Model & Parameter & Unit &  \multicolumn{2}{|c|}{obs1 \& 2} \\
    \hline
         \texttt{xabs} & $N_{\rm H}$ & $10^{20}$\,cm$^{-2}$ & $25\pm4$ & $7\pm3$ \\ 
                        & $\log(\xi)$ & erg cm s$^{-1}$ &\multicolumn{2}{|c|}{$1.7\pm0.2$}  \\
    \hline
        \texttt{relxilld} & q1 & - & $8.0\pm0.3$ & $4.0\pm0.4$ \\
                         & q2 & - &  $2.9\pm0.2$ & $3.1^{+0.7}_{-0.6}$ \\
                         & $R_{\rm b}$ & $r_{\rm g}$ & $3.4^{+0.2}_{-0.4}$ & $5\pm2$ \\
                         & $a_{*}$ & - & \multicolumn{2}{|c|}{>0.97}\\
                         & $i$ & deg  & \multicolumn{2}{|c|}{$32\pm2$} \\
                         & $Z_{\rm Fe}$ & $Z_{\odot}$ & \multicolumn{2}{|c|}{$4.8\pm1.2$} \\
                         & $\log(n_{\rm e})$ & cm$^{-3}$ & \multicolumn{2}{|c|}{$<15.7$} \\
                             & $\log(\xi)$ & erg cm s$^{-1}$ & $1.20\pm0.15$ & $1.3^{+0.3}_{-0.2}$\\
                         & $\Gamma$ & -  & $2.26\pm0.02$ & $2.49\pm0.06$ \\
                         & $f_{\rm refl}$ & - & $10\pm4$ & $3.0\pm1.5$ \\
                         & $\log(F_{\rm X})$ & \ergs & $-11.105\pm0.008$ & $-10.865\pm0.015$\\
    \hline
        & $\chi^{2}/\nu$ & - & \multicolumn{2}{|c|}{867.37/721} \\
        \hline\hline
    \end{tabular}
    \caption{Best-fit parameters obtained by using Model 1.  $F_{\rm X}$ is the unabsorbed flux of the model in the 0.3--10\,keV band. }
    \label{tab_ref}
\end{table*}

The best-fit \red{Model 1} of \src\ suggests that the column density of the warm absorber decreases from $2.5\pm0.4\times10^{21}$\,cm$^{-3}$ in obs1 to $1.0\pm0.3\times10^{21}$\,cm$^{-3}$ in obs2 while its ionisation state remains consistent. We investigate whether it is possible to explain the X-ray spectral variability of \src\ by varying only the intrinsic continuum emission. By linking the column density parameters of two observations, the fit is significant worse below 1~keV with $\chi^{2}/\nu=$\red{910.63/723}. Therefore, although the X-ray variability is dominated by the variable intrinsic continuum emission in \red{Model 1}, the contribution of the variable line-of-sight column density of the warm absorber in \src\ cannot be ignored.

\subsection{Partially-Covering Absorption Model (\red{Model 2})}

\subsubsection{Model set-Up}

In this section, we \red{consider a model based on multiple partially-covering absorbers} \citep[e.g.][]{tanaka04}. In this model, the residuals between 4--8\,keV as shown in Fig.\,\ref{pic_pl} are explained by the Fe~K absorption edge of two high-$N_{\rm H}$ absorbers \citep[e.g.][]{waddell19}. 

\begin{figure}
    \centering
    \includegraphics[width=\columnwidth]{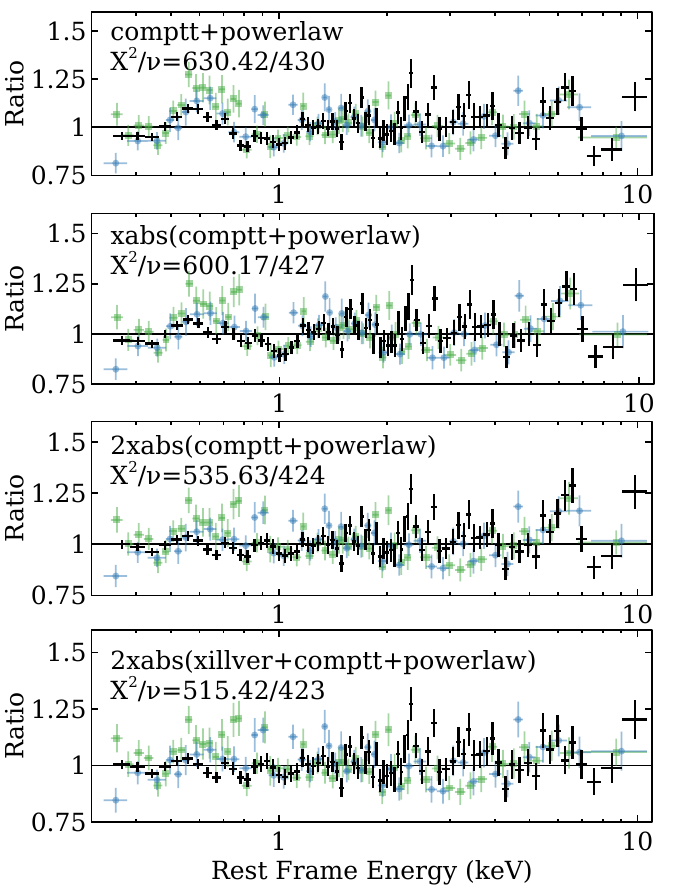}
    \caption{Data/model ratio plots of the obs1 spectra of \src\ fit to different models (red: pn; blue squares: MOS1; green circles: MOS2). See text for more details.}
    \label{pic_abs_step}
\end{figure}

We first fit the soft excess emission of \src\ to a soft Comptonisation model, similar to the model for another NLS1 Mrk~335 in \citet{gallo15}. The \texttt{comptt} model  \citep{titarchuk94,marshall03} is used for this purpose. The combination of warm and hot corona models provides a fit to the obs1 data, for example, with $\chi^{2}/\nu=$\red{630.42/430}. See the first panel of Fig.\,\ref{pic_abs_step} for the corresponding data/model ratio plot. Residuals are seen at 6.4\,keV and <1\,keV, suggesting the existence of Fe~K$\alpha$ emission and low-ionisation absorption. 

Similar to the absorption model in \citet{tanaka04,gallo15}, we consider a low-ionisation partially-covering model to fit the negative residuals at 7\,keV and the absorption features below 1\,keV. The same \texttt{xabs} model as in \red{Model 1} is used to model the photoionisation absorption in the data. 

One additional \texttt{xabs} component improves the fit by $\Delta\chi^{2}=$\red{30} with three more free parameters ($N_{\rm H}$, $\log(\xi)$ and $f_{\rm cov}$). See the second panel of Fig.\,\ref{pic_abs_step} for the corresponding data/model ratio plot. The fit of Fe UTA at 0.8 keV is improved by adding \texttt{xabs}. To further improve the fit below 1\,keV, we add a second \texttt{xabs} model, which decreases $\chi^{2}$ by \red{65} with three more free parameters. 

We then add an additional \texttt{xillver} model \citep{garcia10} with the ionisation parameter fixed at $\log(\xi)=0$ to account for narrow Fe~K$\alpha$ emission from a distant reflector. The additional reflection component improves the fit by $\Delta\chi^{2}=$\red{20} with one more free parameter. The final best-fit model is \texttt{tbnew * xabs1 * xabs2 * (comptt + powerlaw + xillver)} (\red{Model 2}) in XSPEC notations. Similar conclusions are found for obs2. In summary, \red{Model 2} needs two layers of low-ionisation absorption in combination with the warm corona model. 

Note that some positive residuals are still seen at 6\,keV of obs1 using \red{Model 2} (see the last panel of Fig.\,\ref{pic_abs_step}). The combination of narrow Fe~K$\alpha$ emission from the distant reflection model and the Fe~K absorption from partially-covering absorbers is unable to fit the spectra in the iron emission band perfectly. Relativistic correction for the reflection component is still required, although \red{Model 2} is able to provide an acceptable fit to the X-ray continuum emission of \src. In Section\,\ref{hybrid}, we will further improve the fit in the iron emission band by considering a hybrid model including both absorption and disc reflection.

Lastly, \red{Model 1} offers obs1 a better fit than \red{Model 2} with $\Delta\chi^{2}=$\red{9}. The difference of two fits are in not only the iron emission band as described above but also the 8--10\,keV band in the observed frame. Positive residuals are seen when fitting the spectra of obs1 with \red{Model 2}. In comparison, \red{Model 1} fits the spectra better near the upper limit of the EPIC energy range. Future hard X-ray observations, e.g. from \nustar\ \citep{harrison13} or \textit{HEX-P} \citep{madsen19}, may help distinguishing two models in the >10\,keV band.

\subsubsection{Multi-epoch analysis} \label{abs_multi}

By applying \red{Model 2} to two observations, we also obtain reasonably good fits for the continuum emission. The best-fit parameters are shown in Table \,\ref{tab_abs}. \red{By doing so, we find a good fit for both observations with $\chi^{2}/\nu=880.46/721$. Best-fit parameters are shown in Table\,\ref{tab_abs} and the best-fit models are shown in the upper panel of Fig.\,\ref{pic_abs}. Corresponding data/model ratio plots are shown in the lower panels of Fig.\,\ref{pic_abs}.}

In \red{Model 2}, the ionisation states of the two absorbers are consistent in two epochs: the first absorber has an ionisation state of $\log(\xi)\approx2.5$; the second absorber has an ionisation state of $\log(\xi)\approx0.7$.  The covering factor of the first absorber decreases from 0.5 during obs1 to 0.3 during obs2; the same parameter of the second absorber decreases from 0.76 during obs1 to 0.6 during obs2. Furthermore, the optical depth of the warm corona remains consistent with 20 while its temperature increases from \red{0.15}\,keV to \red{0.26}\,keV.

\subsubsection{Results}

In \red{Model 2}, the soft X-ray variability of \src\ is explained by variable line-of-sight absorption and intrinsic continuum emission including the soft excess emission and the hot coronal emission. 

Two partially-covering absorbers in a low-ionisation state, \texttt{xabs1} and \texttt{xabs2},  are needed in \red{Model 2}. The first absorber \texttt{xabs1} has a higher column density, a higher ionisation state and a lower covering factor than \texttt{xabs2}. For instance, \texttt{xabs1} has a column density of \red{2.8}$\times10^{23}$\,cm$^{-2}$, which is approximately \red{18} times the column density of \texttt{xabs2}. The ionisation parameter of \texttt{xabs1} is around \red{64} times the same parameter\footnote{Note that the ionisation parameter is reported in log in Table\,\ref{tab_abs}.} of \texttt{xabs2}. The covering factor of \texttt{xabs1} is lower than that of \texttt{xabs2}. 

The column density and covering factor of the first absorber \texttt{xabs1} increase from the high flux state during obs2 to the low flux state during obs1. The best-fit value of the column density changes by a factor of \red{3}. The covering factor of \texttt{xabs1} increases by a factor of \red{1.6}. The column density of the second absorber \texttt{xabs2} decreases by a factor of 2 from obs2 to obs1 while the covering factor increases by a factor \red{1.3}.

In addition to variable absorption, \red{Model 2} also requires the intrinsic emission to be variable to fit the data. The photon index of the hot coronal emission increases dramatically from \red{1.8} in obs1 to \red{2.5} in obs2. The temperature and the strength of the warm corona also increase from obs1 to obs2. In particular, the unabsorbed flux of the warm coronal emission increases by a factor of 13 in the \xmm\ energy band. 

\begin{figure}
    \centering
    \includegraphics[width=\columnwidth]{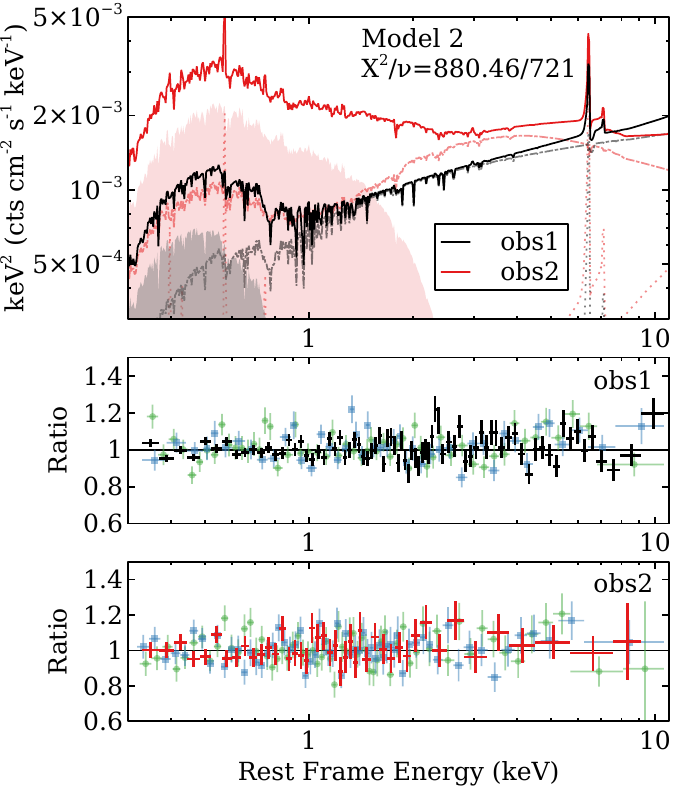}
    \caption{Best-fit models for two \xmm\ observations using Model 2 and corresponding data/model ratio plots (red: obs1 pn; black: obs2 pn; blue squares: MOS1; green circles: MOS2). Solid lines: total models; shaded regions: soft Comptonisation models; dash-dotted lines: power-law components; dotted lines: distant reflectors. Spectra of two observations are fit together. See text for more details.}
    \label{pic_abs}
\end{figure}

We investigate the possibility of fitting the spectra of \src\ using \red{Model 2} without the need of changing the intrinsic emission. The following additional parameters are linked between two epochs: $F_{\rm w}$, $kT$, $F_{\rm h}$, $\Gamma$ and $F_{\rm x}$. By doing so, we obtain a worse fit with $\chi^{2}/\nu=$\red{927.46/726} than the fit presented in Table\,\ref{tab_abs} and Fig.\,\ref{pic_abs} ($\chi^{2}/\nu=$\red{880.46/721}). This model requires intermediate values of the parameters for the linked components: for instance, the flux of the warm corona and the hot corona is $\log(F_{\rm w})=-11.40^{+0.12}_{-0.19}$ and $\log(F_{\rm h})=-10.87^{+0.06}_{-0.07}$; the temperature of the warm corona is $0.21^{+0.03}_{-0.02}$\,keV. Based on the high value of $\chi^{2}$ of this fit, we argue that \red{Model 2} requires both the intrinsic emission and the absorption to be variable to explain the data.

\begin{table*}
    \centering
    \begin{tabular}{ccccccc}
    \hline\hline
         Model & Parameter & Unit & \multicolumn{2}{c}{obs1 \& obs2}\\
    \hline
    \texttt{xabs1} & $N_{\rm H}$ & $10^{22}$\,cm$^{-2}$ & $28^{+9}_{-6}$ & $10\pm7$ \\
                    & $\log(\xi)$ & erg cm s$^{-1}$ & \multicolumn{2}{c}{$2.5\pm0.2$} \\
                    & $f_{\rm cov}$ & - & $0.50\pm0.04$ & $0.31^{+0.17}_{-0.11}$ \\
    \texttt{xabs2} & $N_{\rm H}$ & $10^{22}$\,cm$^{-2}$ &  $1.6\pm0.3$ & $3.4^{+0.7}_{-0.2}$ \\
                    & $\log(\xi)$ & erg cm s$^{-1}$ & \multicolumn{2}{c}{$0.7\pm0.2$} \\
                    & $f_{\rm cov}$ & - & $0.76\pm0.04$ & $0.60\pm0.05$ \\
    \texttt{comptt} & $\tau$ & - & \multicolumn{2}{c}{$20\pm3$} \\
                    & kT & keV & $0.145^{+0.015}_{-0.012}$ & $0.26\pm0.02$ \\
                    & $\log(F_{\rm w})$ & \ergs & $-11.72\pm0.02$ & $-10.6^{+0.3}_{-0.4}$\\
    \texttt{powerlaw} & $\Gamma$ & -  & $1.81\pm0.07$ & $2.50^{+0.13}_{-0.15}$ \\
                      & $\log(F_{\rm h})$ & \ergs & $-11.16^{+0.03}_{-0.02}$ &  $-10.72^{+0.26}_{-0.12}$\\
    \texttt{xillver}  & $\log(F_{\rm x})$ & \ergs &  $-12.49^{+0.15}_{-0.12}$ & $-11.8\pm0.3$ \\
        \hline
        & $\chi^{2}/\nu$ & - & \multicolumn{2}{c}{880.46/721} \\
        \hline\hline
    \end{tabular}
    \caption{Best-fit parameters obtained by using Model 2.}
    \label{tab_abs}
\end{table*}

\subsection{Hybrid Model} \label{hybrid}

In previous two sections, we introduce two spectral models to fit the \xmm\ spectra of \src: one is based on the relativistic disc reflection model (\red{Model 1}) and the other is based on double partially-covering absorption model (\red{Model 2}). The evidence of Fe~K$\alpha$ emission and soft excess in the spectra of \src\ motivates the choice of \red{Model 1}. An additional warm absorption model with a modest column density of 1--$2.5\times10^{21}$\,cm$^{-2}$ is required to fit the Fe UTA at 0.8\,keV. Alternatively, one may fit the soft excess of \src\ with the warm corona model which is included in \red{Model 2}.

In this section, we first compare \red{Model 1} and \red{Model 2} and summarise the interpretations of the soft X-ray variability of \src\ based on two models. We then introduce a hybrid model where the intrinsic emission is described by the relativistic disc reflection model and additional absorption models are used to explain the variability. Such a model can improve the fit in the iron emission and 8--10\,keV bands in comparison to  \red{Model 2}.

\subsubsection{\red{Variable reflection or/and absorption?}}

\src\ shows significant long-term variability in the <2\,keV band on timescales of months and years (see Fig.\,\ref{pic_swift}). We conduct multi-epoch spectral analysis for the two \xmm\ observations of this object based on \red{Model 1} and \red{Model 2} to study the origin of such variability. 

\begin{figure}
    \centering
    \includegraphics{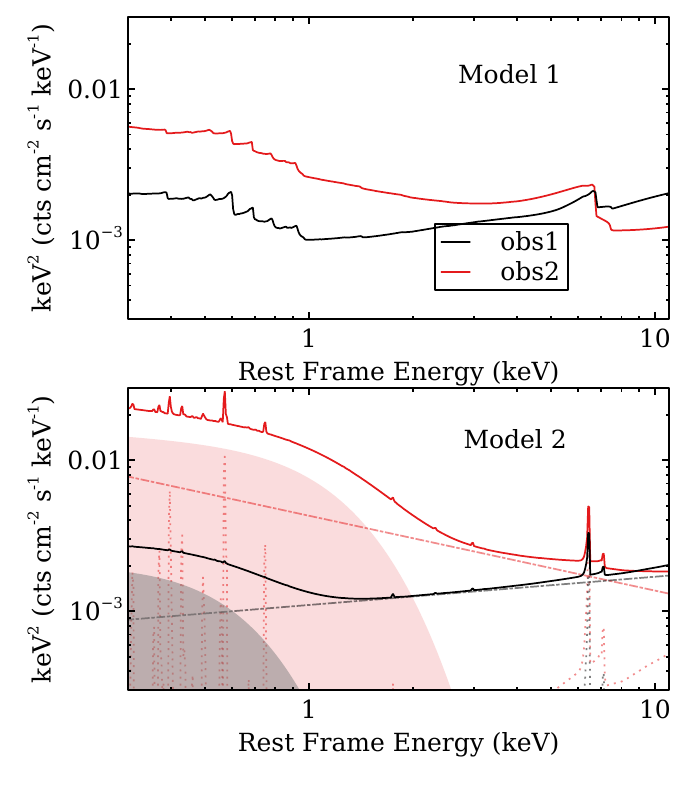}
    \caption{The unabsorbed intrinsic X-ray emission from \src\ during obs1 (black solid lines) and obs2 (red solid lines) suggested by Model 1 (top) and Model 2 (bottom). All low-ionisation absorption in the AGN and Galactic absorption in the foreground are removed in this figure. Dash-dotted lines: hot coronal emission; shaded regions: warm coronal emission; dotted lines: distant reflection components.}
    \label{pic_intrin}
\end{figure}

In \red{Model 1}, the soft X-ray variability is dominated by the intrinsic emission from the hot corona and the reflected emission from the inner accretion disc. The best-fit unabsorbed \red{Model 1} for two \xmm\ observations is shown in the top panel of Fig.\,\ref{pic_intrin}. The X-ray continuum emission is softer during obs2 when the flux is high. The variable disc emissivity profile and reflection fraction of the disc reflection model agrees with the light-bending model \citep{miniutti03}, where the size of the coronal region plays an important role. Variability in the column density of the warm absorption, which shows a consistent ionisation state, contribute to the soft X-ray variability too.

In \red{Model 2}, both absorption and intrinsic emission contribute to the soft X-ray variability of \src. Two low-ionisation partially-covering absorbers are required to fit the broad band spectra. They show a higher covering factor when the observed soft X-ray flux of \src\ is low during obs1. Meanwhile, they remain consistent ionisation states. The unabsorbed intrinsic emission of \red{Model 2} is shown in the lower panel of Fig.\,\ref{pic_intrin}. Emission from both hot and warm corona changes: the hot coronal emission becomes softer and the temperature and strength of the warm corona increase. In comparison with \red{Model 1}, \red{Model 2} requires the photon index of the hot coronal emission to increase by a larger amplitude: $\Gamma$ in \red{Model 1} increases from 2.26 in obs1 to \red{2.49} in obs2; $\Gamma$ in \red{Model 2} increases from \red{1.81} to \red{2.50}. 

Furthermore, \red{Model 1} provides a slightly better fit to the spectra of two observations than \red{Model 2} by $\Delta\chi^{2}=$\red{13} with the same number of parameters. The difference in the goodness of their fits is in the 6--10\,keV band, where the Fe~K emission line is not well fit by \red{Model 2} and some positive residuals are still seen above 8\,keV (see Fig.\,\ref{pic_abs_step} and \ref{pic_warm}).

In addition to \red{Model 2} and \red{Model 1}, we propose a hybrid model where the disc reflection model is used to fit the intrinsic emission and absorption models are still needed to explain the variability. Such a model can improve the fit in the 6--10\,keV band by modelling the broad Fe~K$\alpha$ emission with a relativistic disc model. We also investigate whether the spectral variability can be explained by variable absorption and power-law emission without the need for changes in the size of the corona in this hybrid model.

\subsubsection{Model set-up}

In this section, we present a multi-epoch analysis of the \xmm\ spectra of \src\ based on a hybrid model. The relativistic disc reflection model \texttt{relxilld} is used to model the intrinsic emission. The same model is used in Section \ref{ref_multi}. Two \texttt{xabs} models as in Section \ref{abs_multi} are included to account for absorption. 

In \red{Model 1}, one \texttt{xabs} model is included to fit the Fe UTA of the full-covering warm absorber in \src. Although the variability is dominated by the varying intrinsic emission, the variable column density of the warm absorber cannot be ignored (see Section\,\ref{ref_multi}). In particular, \red{Model 1} also suggests that the size of the coronal region has to change to explain the variable disc emissivity profiles and reflection fractions. In this hybrid model, we study whether it is possible to keep the geometry of the coronal region consistent and interpret the soft X-ray variability with absorption together with changes in the illuminating coronal emission. 

To achieve the goals above, we link most of the parameters in \texttt{relxilld} between two observations. They include the emissivity profile, density, ionisation and reflection fraction of the disc. Besides, the spin of the central BH, the inclination angle and iron abundances of the disc are not expected to change on observable timescales. So, they are linked too. We try to fit the data with linked photon index, but the fit is significantly worse. We, therefore, allow the photon index and normalisation of \texttt{relxilld} to be different in two observations. These two parameters describe the illuminating spectrum of the disc. 

The first absorption model \texttt{xabs1} is required to fit the spectrum of obs1 but not necessary for obs2 when flux is high. We, therefore, link the column density and ionisation parameter of \texttt{xabs1} for two observations. Only an upper limit of the covering factor is found for obs2 ($f_{\rm cov}<0.09$). The second absorption model \texttt{xabs2} remains a consistent ionisation state. The ionisation parameter is thus linked too. We obtain only a lower limit of the covering factor for \texttt{xabs2} at 0.97, suggesting a full-covering geometry. The best-fit ionisation parameter is around 0.9, which is lower than the value of \red{Model 1} but still consistent with typical values in most AGN \citep{reynolds95,laha14}. Similar to \red{Model 1}, the column density of \texttt{xabs2} increases from $6\times10^{20}$\,cm$^{-2}$ during the high flux state (obs2) to $1.1\times10^{21}$\,cm$^{-2}$ during the low flux state (obs1). 

\subsubsection{Results}

\begin{figure}
    \centering
    \includegraphics[width=\columnwidth]{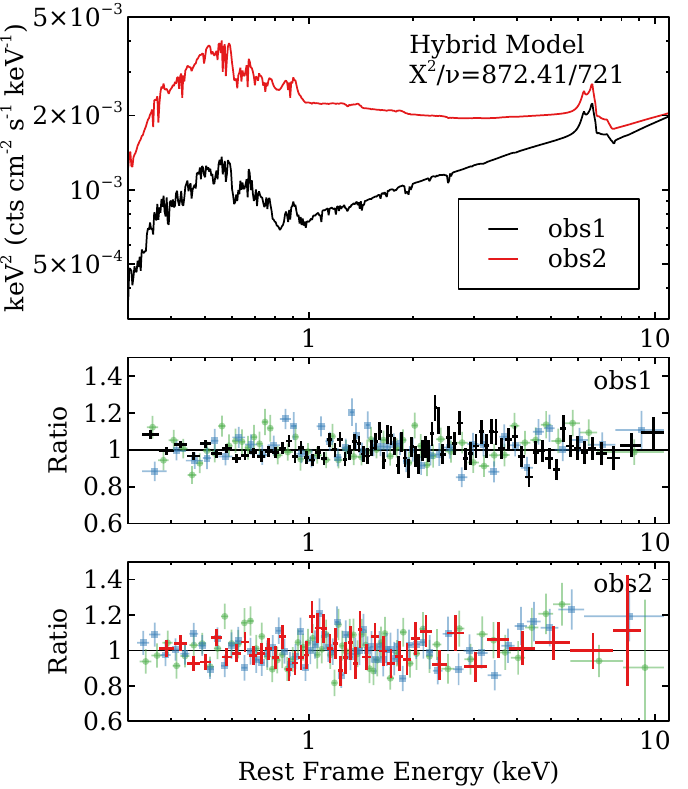}
    \caption{Best-fit models for two \xmm\ observations using a hybrid model and corresponding data/model ratio plots (red: obs1 pn; black: obs2 pn; blue squares: MOS1; green circles: MOS2). Spectra of two observations are fit together. In this scenario, the intrinsic emission is modelled by the relativistic disc reflection model. We assume that the geometry of the coronal region remains consistent. Variable absorption and primary emission from the corona are responsible for the spectral variability.}
    \label{pic_hybrid}
\end{figure}

\begin{table*}
    \centering
    \begin{tabular}{ccccc}
    \hline\hline
         Model & Parameter & Unit & \multicolumn{2}{|c|}{obs1 \& 2} \\
    \hline
         \texttt{xabs1} & $N_{\rm H}$ & $10^{22}$\,cm$^{-2}$ & \multicolumn{2}{c}{$1.7^{+0.7}_{-0.6}$} \\ 
                        & $\log(\xi)$ & erg cm s$^{-1}$ &   \multicolumn{2}{c}{$1.52^{+0.20}_{-0.15}$}\\
                        & $f_{\rm cov}$ & - & $0.47\pm0.07$ & <0.09 \\
    \hline
             \texttt{xabs2} & $N_{\rm H}$ & $10^{21}$\,cm$^{-2}$ & $1.1^{+0.2}_{-0.3}$ & $0.6^{+0.2}_{-0.3}$ \\ 
                        & $\log(\xi)$ & erg cm s$^{-1}$    &\multicolumn{2}{c}{$0.9\pm0.2$}\\
                        & $f_{\rm cov}$ & - & \multicolumn{2}{c}{>0.97} \\
    \hline
        \texttt{relxilld} & q1 & - & \multicolumn{2}{c}{$5.6^{+0.4}_{-0.5}$}  \\
                         & q2 & - & \multicolumn{2}{c}{$2.6\pm0.6$} \\
                         & $R_{\rm b}$ & $r_{\rm g}$ & \multicolumn{2}{c}{$5.5^{+2.0}_{-1.4}$} \\
                         & $a_{*}$ & - & \multicolumn{2}{|c|}{>0.95}\\
                         & $i$ & deg &  \multicolumn{2}{|c|}{$32\pm3$} \\
                         & $Z_{\rm Fe}$ & $Z_{\odot}$ &  \multicolumn{2}{|c|}{$4\pm2$} \\
                         & $\log(n_{\rm e})$ & cm$^{-3}$ &  \multicolumn{2}{|c|}{$16.6^{+0.8}_{-1.0}$} \\
                             & $\log(\xi)$ & erg cm s$^{-1}$ & \multicolumn{2}{c}{$1.06\pm0.02$}\\
                         & $\Gamma$ & - & $2.29\pm0.02$ & $2.51\pm0.03$ \\
                         & $f_{\rm refl}$ & - & \multicolumn{2}{c}{$3.2\pm0.3$} \\
                         & $\log(F_{\rm X})$ & \ergs & $-11.01\pm0.02$ & $-10.82^{+0.03}_{-0.02}$ \\
    \hline
        & $\chi^{2}/\nu$ & - &  \multicolumn{2}{|c|}{872.41/721} \\
        \hline\hline
    \end{tabular}
    \caption{Best-fit parameters obtained by using a hybrid model. $F_{\rm X}$ is the unabsorbed flux of the model in the 0.3--10\,keV band. }
    \label{tab_hybrid}
\end{table*}

The hybrid model introduced above provides a good fit to both observations with $\chi^{2}/\nu=$\red{872.41/721}. The fit is slightly worse than \red{Model 1} by $\Delta\chi^{2}=5$ with the same number of free parameters, and better than \red{Model 2} by $\Delta\chi^{2}=$\red{8} also with the same number of parameters. Best-fit models are shown in Fig.\,\ref{pic_hybrid} and best-fit parameters are shown in Table\,\ref{tab_hybrid}. The fit in the iron emission band is improved compared to \red{Model 2}. Some residuals are still seen above 8\,keV in obs1 but reduced from the fit of \red{Model 2}.

Assuming the geometry of the coronal region remains consistent, we obtain an intermediate value of reflection fraction for the disc (\red{3.2}$\pm0.3$), which lies between the values for two observations inferred by \red{Model 1}. Similar conclusion is found for the emissivity profile of the disc, which is flatter than the one for obs1 in \red{Model 1} but steeper than the one for obs2. Consistent measurements for the spin of the BH, the inclination and iron abundances of the disc are achieved. Tentative evidence shows that a higher disc density is required when fitting two spectra with the same reflection model. The 90\% confidence range of the density parameter is $\log(n_{\rm e}/{\rm cm}^{-3})=16.6^{+0.8}_{-1.0}$. Considering a 3-$\sigma$ uncertainty range\footnote{The hard lower limit of $n_{\rm e}$ in \texttt{relxilld} is $10^{15}$\,cm$^{-3}$.}, we obtain only an upper limit at $10^{18}$\,cm$^{-3}$.

\begin{figure}
    \centering
    \includegraphics{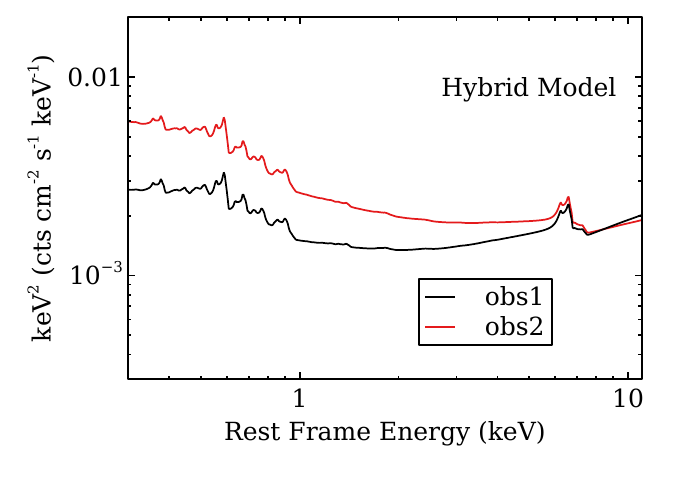}
    \caption{Same as Fig.\,\ref{pic_intrin} but for the hybrid model.}
    \label{pic_intrin_hybrid}
\end{figure}

In this hybrid model, the strength and shape of the coronal emission are allowed to be different in two epochs. The resulting reflection spectrum changes accordingly with a consistent flux fraction. We show the best-fit unabsorbed models in Fig.\,\ref{pic_intrin_hybrid}. The intrinsic emission is softer during obs2 than obs1. The photon index of the power-law emission increases from \red{$2.29$} to \red{$2.51$}. The unabsorbed flux of the \texttt{relxilld}, however, does not change a lot ($\log(F)=$\red{-11.01} in obs1 and  $\log(F)=$\red{-10.82} in obs2). 

Additional variable absorption is needed in the hybrid model. The first absorber \texttt{xabs1} is in a higher ionisation state than the second absorber \texttt{xabs2}. \texttt{xabs1} has a covering factor of 0.47 during obs1 when the soft X-ray flux of \src\ is low. During the high flux state, obs2 requires no \texttt{xabs1}. We estimate the upper limit of its covering factor to be at 0.09. The second absorber is similar to typical warm absorbers seen in  other AGN. The covering factor is consistent with 1 (>0.97). The column density of \texttt{xabs2} decreases by a factor of 2 from obs1 to obs2.

In summary, the hybrid model is also able to explain to the multi-epoch variability of \src. This model provides a slightly worse fit than \red{Model 1} but a better fit than \red{Model 2} with improvements of fitting the data in the 6--10\,keV band. In this hybrid model, the soft X-ray variability of \src\ is explained by variable power-law emission from the corona. The reflected emission from the accretion disc changes accordingly without changing the geometry of the corona. During the low  flux state, a partially-covering absorber of $\log(\xi)=$\red{1.5} crosses our line of sight to the source with a covering factor of around 47\%. During the high soft X-ray flux state, this absorber moves out of our line of sight. Additional full-covering warm absorber similar to those in other AGN is needed to fit the data. The warm absorber shows a slightly higher column density in obs1 than obs2. 

\red{The hybrid model improves the fit in the iron emission band compared to Model 2 by including the relativistic disc reflection model. No additional component is required to fit the soft excess emission. We investigate whether an additional soft Comptonisation component as in Model 2 is able to further improve the fit. The \texttt{comptt} model is used for this purpose. We find that the fit is not significantly improved as the disc reflection component accounts for both the iron emission and the soft excess of \src. We fix the parameters of the \texttt{comptt} model at the best-fit values in Model 2 (see Table\,\ref{tab_abs}) and then obtain an upper limit for the contribution of the soft Comptonisation component in the 0.3--10\,keV band, $F_{\rm w}<1.6\times10^{-13}$~\ergs\ and $<4\times10^{-13}$~\ergs\ respectively for obs1 and obs2.}

\section{Discussion} \label{discuss}

We present different spectral models for the \xmm\ observations of \src. \red{The first model is based on the relativistic disc reflection model and requires a variable size in the coronal region to explain the data. The second model is based on multiple partially-covering absorbers in combination with the warm corona model. The observed soft X-ray variability is interpreted by variable line-of-sight absorption and soft Comptonisation emission. We also propose a third hybrid model, where both multiple absorption and disc reflection models are included. In this scenario, the size of the corona remains consistent while intrinsic variability in its emission is expected. The soft X-ray variability is dominated by one partially-covering absorber in the low X-ray flux state, which leaves our line of sight during the high X-ray flux state.}

\subsection{Variable Disc Reflection (\red{Model 1})}

In \red{Model 1}, the spectral variability is dominated by the Comptonisation emission from the hot corona. The photon index of the coronal continuum emission is higher when the X-ray luminosity is higher, which was often seen in many other Sy1s \citep[e.g.][]{jiang18,wu20}. 

The resulting disc reflection component changes according to the variable illuminating coronal emission. In particular, the lower flux state observation (obs1) has a higher reflection fraction than the higher flux state observation (obs2). This can be explained by the light-bending effects of the variable size of the corona in \src\ \citep[e.g.][]{miniutti03, reis15, jiang18}. 

The reflection fraction parameter in \texttt{relxilld} is defined as the ratio between the intensity of the coronal component that reaches the disc and the one seen by the observer, and the value of this parameter can easily exceed unity when the corona is compact \citep{dauser16}. The best-fit reflection fraction parameters for obs1 and obs2 are respectively \red{10} and \red{3}. Such a large amplitude change of the reflection fraction parameter was also seen in the multi-epoch variability of other AGNs \citep{jiang18d}. The high values of reflection fraction parameters suggest that the corona is within a region of <3\,$r_{\rm g}$ assuming a simple `lamp-post' geometry\footnote{The `lamp-post' model assumes a point-like, isotropic geometry of the corona located on the rotating axis of the BH.} \citep{dauser16}. The flatter disc emissivity profile also suggests the existence of a more extended coronal region in \src\ in obs2 than obs1. Because the outer region of the accretion disc is more illuminated when the coronal region is larger \citep[e.g.][]{gonzalez17}. \red{The corona is known to be very compact  within a few gravitational radii in many AGN as well as \src\ according to advanced timing analyses in the X-ray band \citep[e.g.][]{fabian09,reis13}. Similar conclusions were found in microlensing events of AGN \citep{morgan08,chartas17}. The compact corona also agrees with the predictions of some coronal models, such as magnetic reconnection where the magnetic field increases strength towards smaller radii \citep{merloni01}, the base of the jet \citep{ghisellini04} and pair productions in the magnetosphere around the central BH \citep{hirotani98,chen20}. }


We estimate the distance of the warm absorber in \red{Model 1} from the central BH using the best-fit ionisation parameter $\xi=$\red{50}\,erg\,cm\,s$^{-1}$. The bolometric luminosity is estimated to be $9\times L_{\rm 5100}\approx3.6\times10^{45}$\,erg\,s$^{-1}$ \citep{peterson04}. Assuming a density of $n=10^{9}$\,cm$^{-3}$ \citep{reynolds95} and an isotropic illuminating source, we estimate the location of the warm absorber in \src\ to be around $d=\sqrt{\frac{L_{\rm Bol}}{4\pi n \xi}}\approx\red{6}\times10^{17}$\,cm.

In our calculation, the assumption of $n=10^{9}$\,cm$^{-3}$ for the warm absorber is based on the agreement of recombination timescale and the variability timescale of the illuminating emission in MCG-6-30-15 \citep{reynolds95}. If the recombination timescale is larger than the variability timescale, photoionization equilibrium would not apply. The primary emission of MCG-6-30-15 is much more variable than \src\ on timescales of kiloseconds. So, the density of the warm absorber in \src\ is allowed to be lower than $n=10^{9}$\,cm$^{-3}$ as the recombination timescale is approximately proportional to $n^{-1}$ \citep{reynolds95}. Furthermore, we also assume an isotropic illuminating source and apply a factor of $4\pi$ in the calculation, which can also be lower. So, the estimated distance of $\red{6}\times10^{17}$\,cm is only the lower limit.

Lastly, \red{Model 1} also provides an estimation of the BH spin of \src. By fitting two epochs simultaneously and linking their spin parameters, we achieve a high BH spin of $a_{*}>0.97$ in Section\,\ref{ref_multi}. Previously, \citet{reynolds14} noted the tentative evidence that the most massive black holes ($M_{\rm BH}>10^8M_{\odot}$) and the least massive black holes ($M_{\rm BH}<10^6M_{\odot}$) may have more modest spins by compiling a number of measurements in previous work. The properties of the host galaxy may play an important role in the evolution of the BH spin \citep{senana14}. \src\ may host one of the few massive BHs \citep{peterson04,ho08} which have a high BH spin. Future reflection studies of AGN with similar high BH masses enable us to better understand the spin distribution in a wider range of BH masses. 

\subsection{Variable Absorption and Intrinsic Soft X-ray Emission (\red{Model 2})}

\red{Model 2} provides multi-epoch spectra of \src\ a slightly worse fit than \red{Model 1} by $\Delta\chi^{2}=\red{13}$ with the same number of parameters. In \red{Model 2}, the ionisation states and the column densities of two absorbers are also consistent within their 90\% confidence ranges between two observations. The soft X-ray variability is mainly caused by the variable covering factors of the absorbers: the covering factor of the first absorber increases from 0.34 in obs2 to 0.50 in obs1. We estimate the location of the absorbers according to the X-ray variability assuming the absorbers are orbiting around the central BH: the observed soft X-ray flux of \src\ increases from $\log(F)=-11.47$ in the 0.3--3\,keV band in 2003 to $\log(F)=-11.14$ in 2015 (see Table \ref{tab_obs}) by a factor of more than 2 in 12 years. Meanwhile, the X-ray flux of \src\ does not show large-amplitude variability on kilosecond timescales as many other NLS1s. 
 
We extract long-term lightcurves of \src\ based on \swift\ observations in the archive to investigate how rapidly the soft X-ray emission of \src\ varies. The first two panels of Fig.\,\ref{pic_swift} show the X-ray lightcurves. The X-ray flux of \src\ can amplify by a factor of up to 5 during this long-term period of \swift\ observations. The fifth and sixth \swift\ observations are separated by 6 months in 2010. The soft X-ray flux changes by a factor of 2.7. Similar large-amplitude variability on timescales of months is also seen in other intervals.  Assuming a BH mass of $4\times10^{8}M_{\odot}$, the orbital period at $r=6.8\times10^{13}$\,m$\approx100r_{\rm g}$ is approximately 6 months. So, if the soft X-ray variability observed by \swift\ results from variable absorption only as inferred by \red{Model 2}, the absorbers need to be located less than $\approx100r_{\rm g}$ from the central BH of \src\ to explain the observed large-amplitude X-ray variability on timescales of months. 

In addition to the variability in absorption, dramatic changes in the intrinsic emission are also required by \red{Model 2} to explain the spectral variability. The unabsorbed flux of the warm coronal emission increases by a factor of 13 in the \xmm\ energy range. The temperature of the corona increases from 0.18 keV to 0.23 keV. 

\red{In the warm corona model, the soft Comptonisation emission from the warm corona is often found to dominate the extreme-UV band \citep[e.g.][]{jin17}. We extend our best-fit models to the extreme-UV band and calculate their predicted flux in the 0.01--0.1~keV band where the soft Comptonisation emission peaks. The warm corona model suggests that the unabsorbed extreme-UV flux of the non-thermal emission in \src\ increase by a factor of 9 from $4\times10^{-12}$ \ergs\ to $3.6\times10^{-11}$ \ergs. A significant change in the bolometric luminosity is thus expected in Model 2. It is interesting to note that the Eddington ratio of \src\ is estimated to be around 5\% based on the observed 5100\AA\ luminosity (see Section\,\ref{intro}). The soft excess emission plays an important role in the estimation of bolometric luminosities for AGN at a few percent of Eddington when the warm corona model is used \citep{noda18}.}

\red{In comparison, the relativistic disc reflection model in Model 1 suggests that the unabsorbed flux of the non-thermal emission in \src\ increases by a factor of only 2.9 from $2\times10^{-12}$ \ergs\ to $5.7\times10^{-12}$ \ergs\ in the 0.01--0.1\,keV band. A much smaller increase in the extreme-UV luminosity is required in the disc reflection model. Unfortunately, due to Galactic absorption, it is not possible to measure the luminosity of \src\ in this energy band and test each model.}

\red{The Optical Monitor on \xmm\ provides complementary optical and UV views of \src\ at longer wavelengths, although we are unable to measure the extreme-UV flux of this object. The observed magnitude of \src\ increases by a factor of 2 from $13.597\pm0.008$ during obs1 to $13.307\pm0.007$ during obs2 in the UVW1 band and a factor of 2.3 from $13.692\pm0.008$ to $13.337\pm0.007$ in the UVM2 band. Photometric observations with only two filters are not able to constrain the thermal emission from the disc and thus the spectral energy distribution of this object. But we find that the observed UV flux at longer wavelengths varies by a similar factor as the extreme-UV flux predicted by the disc reflection model rather than the warm corona model.}

\subsection{Hybrid Model}

Based on the fits of \red{Model 1} and \red{Model 2}, we propose for a hybrid model, where the intrinsic emission is modelled by disc reflection and absorption is still required to explain the spectral variability in \src. This hybrid model also provides a good fit to the data. Such a model improves the fit in the iron emission band in comparison with the absorption model, \red{Model 2}.

In the hybrid model, the variability of the intrinsic X-ray emission from \src\ is caused by the variable intrinsic power-law emission from the corona. The flux of the coronal emission increases by a small amplitude from $9.5\times10^{-12}$\,\ergs\ to $1.4\times10^{-11}$\,\ergs\ while its photon index increases from $2.29$ to $2.51$. We assume the coronal region remains the same geometry during the two observations in this model. The resulting disc reflection spectrum changes according to the variable power-law emission with a consistent reflection fraction.

Furthermore, two layers of absorption are needed to explain the spectral variability of \src: one full-covering warm absorption with a modest column density of approximately $10^{21}$\,cm$^{-3}$ and one partially-covering absorber in a higher ionisation state of $\log(\xi)\approx1.7$. The partially-covering absorber has a covering factor of 47\% in obs1 when the flux is low and disappear on the line of sight during obs2. Assuming this partially-covering absorber contributes to the variability of \src\ on timescales of months observed by \swift, this absorber needs to be located in the region of $<100$\,$r_{\rm g}$. Variable absorption at a large distance along the line of sight has been seen in other AGN too \citep[e.g.][]{grupe04c,parker14b,kaastra18,miller21}. Objects like NGC~6814 require both disc reflection and absorption models to fit the X-ray data \citep{gallo21}. Variable absorption at a large distance from the innermost X-ray emission region plays an important role in the observed variability of these sources.

\section{Conclusion} \label{conclude}

In this work, we investigate the nature of the soft X-ray variability of the NLS1 AGN \src\ based on two \xmm\ observations in the archive. The soft excess emission of \src\ shows a very different spectral shape in these two observations. We apply two models to the EPIC data of the source, one based on relativistic disc reflection (\red{Model 1}) and the other based on partially-covering absorption in combination with the warm corona model (\red{Model 2}). 

In the reflection scenario, the X-ray variability of \src\ is dominated by the variable emission from the hot corona. The disc reflection component changes accordingly. The anti-correlation between the reflection fraction parameter and the X-ray flux suggests a variable coronal geometry in \src. The flatter disc emissivity profile also supports the conclusion that the coronal region of \src\ is more extended during the high X-ray flux state than the low X-ray flux state. The variable modest column density of the line-of-sight warm absorption also contributes to the soft X-ray variability.

In the absorption scenario,  the two high-$N_{\rm H}$ absorbers produce strong Fe~K edges in the Fe~K band. An additional low-ionisation reflection component is required to fit the Fe~K$\alpha$ emission of \src. The variability of \src\ is caused by the variable covering factor of the line-of-sight absorbers in this model, while they remain in a consistent low-ionisation state. \swift\ observations suggest that the absorbers have to be located within a region of $r<\approx100r_{g}$ to explain the large-amplitude X-ray variability of \src\ on timescales of months. In addition, the intrinsic emission from the AGN also needs to be variable: the temperature and strength of the warm corona increases. 

To further improve the fit in the iron band based on \red{Model 2}, we investigate the possibility of a hybrid model, where 1) we assume that the geometry of the coronal region remains the same during the two observations; 2) the intrinsic power-law emission from the corona is allowed to vary; 3) variable absorption is used to explain the soft X-ray variability. Such a hybrid model offers a slightly better fit than \red{Model 2}. The variable absorbers in the hybrid model have two component, one full-covering warm absorber as those commonly seen in typical AGN \citep{reynolds95} and one partially-covering absorber. The partially-covering absorber has a covering factor of 47\% during obs1 when the observed soft X-ray flux of \src\ is low and completely moves out of our line of sight during obs2 when the observed flux is high.

\red{Model 1, 2} and the hybrid model provide CCD-resolution data of \src\ with a similarly good fit. However, a different number of absorbers in different ionisation states is needed in them.
Unfortunately,  archival EPIC data of \xmm\ do not have high enough spectral resolutions to distinguish these models. The RGS observation of obs2 is off-axis. No high-resolution soft X-ray spectrum during obs2 is available for comparison. But future high-resolution X-ray observations, e.g. from \athena\ \citep{barcons17} and \textit{XRISM} \citep{xrism20}, might provide us with a unique opportunity to constrain any of the models for \src\ by resolving multiple spectral components, \red{e.g., from disc reflection \citep{parker22b} and warm absorption \citep{parker22a}}.

\section*{Acknowledgements}

This paper was written during the worldwide COVID-19 pandemic in 2022. We acknowledge the hard work of all the health care workers around the world. We would not be able to finish this paper without their protection. J.J. acknowledges support from the Leverhulme Trust, the Isaac Newton Trust and St Edmund's College, University of Cambridge. \red{This work is based on observations obtained with \xmm, an ESA science mission
with instruments and contributions directly funded by ESA Member States and NASA. This project has made use of
the Science Analysis Software (SAS), an extensive suite to process the data collected by the XMM-Newton observatory.}

\section*{Data Availability}

All the data can be downloaded from the HEASARC website at https://heasarc.gsfc.nasa.gov. \red{The \texttt{relxill} package can be downloaded at https://www.sternwarte.uni-erlangen.de/dauser/research/relxill/index.html.}






\bibliographystyle{mnras}
\bibliography{ugc} 





\bsp	
\label{lastpage}
\end{document}